\documentclass{article}
\usepackage{dcase2018,amsmath,graphicx,url,times,booktabs, tabularx}
\usepackage{color}
\usepackage{dblfloatfix}
\title{A multi-device dataset for urban acoustic scene classification}
\name{Annamaria Mesaros,
      Toni Heittola,
      Tuomas Virtanen$ \sthanks{This work has received funding from the European Research Council under the ERC Grant Agreement 637422 \mbox{EVERYSOUND}.}$
      }

\address{Tampere University of Technology, Laboratory of Signal Processing, Tampere, Finland\\ \{annamaria.mesaros, toni.heittola, tuomas.virtanen\}@tut.fi
}

\begin{document}

\ninept
\maketitle

\begin{sloppy}

\begin{abstract}
This paper introduces the acoustic scene classification task of DCASE 2018 Challenge and the TUT Urban Acoustic Scenes 2018 dataset provided for the task, and evaluates the performance of a baseline system in the task. As in previous years of the challenge, the task is defined for classification of short audio samples into one of predefined acoustic scene classes, using a supervised, closed-set classification setup. 
The newly recorded TUT Urban Acoustic Scenes 2018 dataset consists of ten different acoustic scenes and was recorded in six large European cities, therefore it has a higher acoustic variability than the previous datasets used for this task, and in addition to high-quality binaural recordings, it also includes data recorded with mobile devices. 
We also present the baseline system consisting of a convolutional neural network and its performance in the subtasks using the recommended cross-validation setup. 
\end{abstract}

\begin{keywords}
Acoustic scene classification, DCASE challenge, public datasets, multi-device data
\end{keywords}

\section{Introduction}
\label{sec:intro}
\vspace{-4pt}

%this part about dcase in general and about task 1 definition and all three subtasks shortly 

%DCASE Challenge is an international evaluation campaign focusing on topics related to environmental audio classification, being currently at its fourth edition.
Acoustic Scene Classification is a regular task in the Detection and Classification of Acoustic Scenes and Events (DCASE) challenge series, being present in each of its editions up until now. The standard setup of the task as a basic multiclass classification problem makes the task easily approachable also for the beginner in this field, resulting in large number of participants in the previous DCASE challenges. 
In the first three editions of the challenge, the acoustic scene classification task has received the highest number of submissions among the available tasks, with 17 submissions in 2013 \cite{Stowell:2015}, 48 submissions in 2016 \cite{Mesaros2018_TASLP}, and 97 submissions in 2017 \cite{mesaros-iwaenc2018}. 
% 17 submissions, 12 teams in task 1 dcase 2013

% At first, acoustic scene classification approaches were dominated by classical machine learning methods such as Gaussian mixture models (GMM) [cite], support vector machines (SVM) [cite], non-negative matrix factorization (NMF) \cite{Bisot2017}. From the second edition, the best ranked methods have used either ensemble methods \cite{Eghbal-Zadeh2017,Marchi2016}, deep learning \cite{Valenti2017,Bae2016} or special techniques involving generative adversarial networks in combination with deep learning \cite{Mun2017} and various deep learning architectures [cite few from 2017]. 
%Besides the general popularity of deep learning based methods, the development of such methods in the task was facilitated by the larger datasets provided within the challenge each year.
%Detailed analysis of the submissions from previous editions are available in  \cite{Barchiesi2015,Mesaros2018_TASLP,mesaros-iwaenc2018}.

Each consecutive edition of the challenge has brought a new and larger dataset than previous edition, facilitating use of recent machine learning techniques using deep neural networks that rely on large amounts of data for training. In 2013, the acoustic scene classification task used a development dataset consisting of 10 acoustic scenes each with 10 examples of 30~s, and an evaluation dataset of the same size \cite{Stowell:2015,Barchiesi2015}.  In 2016, 15 scene classes were used, each with 78 examples of of 30~s in the development set, and 26 examples per class in the evaluation set \cite{Mesaros2018_TASLP}. This dataset offered higher acoustic variability than before through its higher number of classes, recording locations and amount of data, and it was the first suitable for use of deep learning methods. 

In DCASE 2017, the acoustic scene classification task was made more difficult by using 10~s audio segments, by re-segmenting the complete data available in 2016 (both development and evaluation sets), having 312 segments of 10~s per scene class. A new evaluation dataset was recorded in similar locations approximately one year later than the development data, containing 108 segments of 10~s per class. The temporal gap between the recordings created an unexpected mismatch in acoustic conditions, causing a significant drop in performance in all systems between development and evaluation sets \cite{mesaros-iwaenc2018}. 
Outside of DCASE challenge, there are only few other publicly available datasets for acoustic scene classification, notably the LITIS dataset \cite{Rakotomamonjy2015}, containing 19 classes and having approximately 25 hours of audio, recorded using a mobile phone; the Defreville-Aucouturier environmental audio dataset \cite{Aucouturier2007} with 4 main classes (11 detailed classes) and approximately 4 hours of audio; and the UEA Environmental noise datasets \cite{Ma2003} with 10 classes and approximately 4 hours of audio in 2 series recorded with different devices. Of these, only the LITIS dataset has an adequate size for modern machine learning methods. 

DCASE 2018 challenge introduces a new dataset for acoustic scene classification, having a number of ten classes and 24 hours of high-quality audio. It has smaller number of classes than data from previous challenges, but it is much larger in size and acoustic variability, having been recorded in multiple cities across Europe. This is the largest freely available dataset to date, comparable in size to the LITIS dataset, but it is the only one having recordings in multiple countries, while all other publicly available datasets (within and outside of DCASE) are recorded within a single country or city.

At the same time, parallel recordings performed with different devices provide additional variability in the channel properties, allowing an additional subtask for studying the classification problem in mismatched conditions. All previous public evaluations have been done in matched conditions, with a single device used for recording all data, including evaluation data, but in actual usage scenarios of the methods, channel mismatch could be encountered through device mismatch or difference in recording conditions. Other publicly available datasets contain audio recorded with only one type of device, with small exceptions (e.g. \cite{Ma2003}) that do not permit a large-scale study of mismatched devices.
A mismatch usually causes a large drop in performance of machine learning based systems, as noticed in DCASE 2017, therefore this new dataset allows development of techniques that can cope with the mismatch. 

This paper presents the subtasks and dataset used for Task 1 in DCASE 2018. Section \ref{sec:recording-proc} presents the data recording procedure. Section \ref{sec:task-def} introduces the task definition and specific details on the subtasks, while Section \ref{sec:experim-setup} gives details on the experimental setup, including database statistics for each subtask. Section \ref{sec:results} presents the baseline system architecture and the results obtained on the provided experimental setup, and Section \ref{sec:concl} presents conclusions and future work. 
\vspace{-8pt}

\section{Data recording procedure}
\label{sec:recording-proc}
\vspace{-4pt}

The TUT Urban Acoustic Scenes 2018 dataset was collected during February-March 2018, containing recordings of ten acoustic scenes, recorded in six large European cities: Barcelona, Helsinki, London, Paris, Stockholm, and Vienna. Acoustic scenes included are: airport, shopping mall (indoor), metro station (platform, underground), pedestrian street, public square, street (medium level of traffic), traveling by tram, bus and metro (underground), and urban park. Each scene class was defined beforehand, and suitable locations were selected based on the description.

For each city and each scene class, multiple different locations were used to record audio, i.e. different streets, different metro stations, etc. For each such location there are 5-6 minutes of audio, recorded in 2-3 sessions of few minutes each, with a small temporal gap between them. The original recordings were split into segments with a length of 10 seconds that are provided in individual files. Recording locations are numbered and used to identify all audio material from the same location when partitioning the dataset for training and testing. The information available in the dataset consists of: acoustic scene class, city, and recording location IDs. 

Recordings were made using four devices that captured audio simultaneously.
The main recording device consists in a Soundman OKM II Klassik/studio A3, electret binaural in-ear microphone and a Zoom F8 audio recorder using 48~kHz sampling rate and 24 bit resolution. The microphones were worn in the ears, therefore the recorded audio mimics the sound that reaches the human auditory system of the person wearing the equipment. This equipment is further referred to as device A. 

At the same time, the audio was captured using three other mobile devices (e.g. smartphones, cameras), resulting in audio recordings of different quality. We further refer to these devices as B, C, and D. All simultaneous recordings are time synchronized using Panako acoustic fingerprinting system \cite{six2014panako}.
% info about devices which should not be available in the first public draft: 
%----------------------------------
The used mobile devices are the following: device B is a Samsung Galaxy S7, device C is IPhone SE, and device D is a GoPro Hero5 Session. During recording, the Samsung phone was handheld at torso height, the IPhone was worn in a sleeve attached to the strap of a backpack, while the GoPro was mounted on the other strap. The mobile phones recorded single channel audio with a sampling frequency of 44.1~kHz, while the audio recorded by the GoPro is originally stereo, compressed,  sampled at 48~kHz. 
% ------------------------------------ until here 

Two different versions of the dataset were provided for system development, namely TUT Urban Acoustic Scenes 2018, containing only material recorded with device A, and TUT Urban Acoustic Scenes 2018 Mobile, containing material recorded with devices A, B and C. 
All datasets are freely available.
% \footnote{ \\
% TUT Urban Acoustic Scenes 2018, Development dataset: \\ https://doi.org/10.5281/zenodo.1228142\\
% TUT Urban Acoustic Scenes 2018, Evaluation dataset: \\
% https://doi.org/10.5281/zenodo.1293883\\
% TUT Urban Acoustic Scenes 2018 Mobile, Development dataset:\\
% https://doi.org/10.5281/zenodo.1228235\\
% TUT Urban Acoustic Scenes 2018 Mobile, Evaluation dataset:\\
% https://doi.org/10.5281/zenodo.1293901
% }
\footnote{
TUT Urban Acoustic Scenes 2018: \\ 
\scriptsize{
https://doi.org/10.5281/zenodo.1228142,
https://doi.org/10.5281/zenodo.1293883
}
}
\footnote{
TUT Urban Acoustic Scenes 2018 Mobile:\\
\scriptsize{
https://doi.org/10.5281/zenodo.1228235, 
https://doi.org/10.5281/zenodo.1293901
}
}

\section{Task definition}
\label{sec:task-def}
\vspace{-4pt}

Acoustic scene classification is defined as labeling one audio sample as belonging to one of predefined classes associated to acoustic scenes. There are labeled example training data available for all the classes, and therefore the task is an example of a supervised classification problem, having a closed set of categories, as illustrated in Fig. \ref{fig:task1}.
In the DCASE 2018 acoustic scene classification task there are three subtasks, offering more variety and degree of difficulty for the task, and at the same time extending the basic task towards real-life applications where there may be mismatch between training and evaluation data recording devices, or there may be different sources of training data available. 

\begin{figure}
    \centering
    \includegraphics[width=0.75\linewidth]{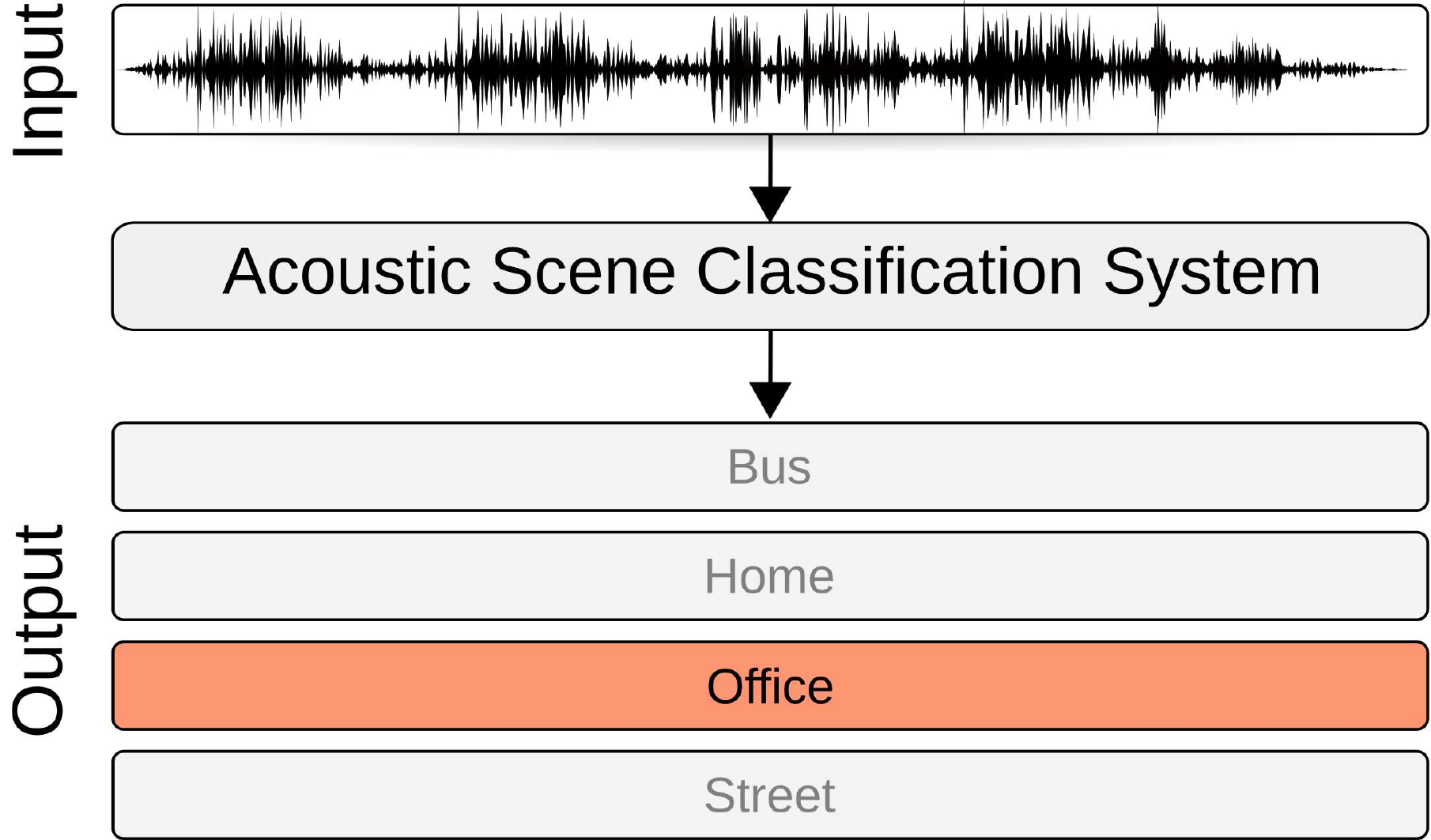}
    \caption{Acoustic scene classification example}
    \label{fig:task1}
    \vspace{-10pt}
\end{figure}

\textbf{Subtask A}: \textit{Acoustic Scene Classification} is the typical acoustic scene classification task as encountered previously, where all data, both development and evaluation, are recorded with a high quality device. In this subtask there is no mismatch in recording conditions besides the natural variation of weather, people at the scene, etc, which are not under control of the recorder, but are natural manifestations of the recorded scenes. 

\textbf{Subtask B}: \textit{Acoustic Scene Classification with mismatched recording devices} illustrates the problem of creating a system that could be used  with multiple devices that record audio of varying quality. In this subtask there is mismatch in audio channels between the development and evaluation sets, which must be accounted for in the system: the training data was recorded with a device providing high-quality audio, while the evaluation data was recorded with multiple devices, resulting therefore in mismatched audio channel. Some amount of parallel data, which was recorded simultaneously with three devices, was also available for training.

\textbf{Subtask C} breaks away from the previous challenge rules against external data sources and allows use of external data and transfer learning to solve the problem provided in subtask A. In subtasks A and B all the participants have the same data available for system development, putting all participants to an equal starting point. In subtask C, systems may be relying on various sources of data, in any form, to study the possible improvement provided by using more data. As a rule for the challenge, the participants were required to use only external datasets that are publicly available for free, and they were also required to inform the organizers about the external data sources for maintaining a list of such resources on the challenge website.

\section{Experimental setup}
\label{sec:experim-setup}
\vspace{-4pt}

The experimental setup is similar for the three subtasks, with the same basic classification problem framed in different ways. In subtask A, only data from device A (high-quality audio) is used, while for subtask B, some data from devices B and C is available as parallel recordings. Subtask C allows the use of external data and transfer learning, but does not provide any additional data, only indicates some sources of data that could be used.
For each subtask, a development set was provided, together with a training/test partitioning for system development. Participants were required to report performance of their system using this train/test setup in order to allow comparison of systems on the development set. 

The total amount of recorded audio was partitioned into development and evaluation subsets, each containing data from all cities and all acoustic scenes. The development dataset was published when opening the task, provided with full metadata information. The evaluation dataset was published as audio only; the metadata of this part is kept secret, and the evaluation of systems is performed by the task organizers, based on the predicted scene labels that participants have to submit when participating the challenge.

\subsection{Development datasets}

\textbf{TUT Urban Acoustic Scenes 2018} development dataset consists of recordings from all six cities, having 864 segments for each acoustic scene (144 minutes of audio). The total size of the dataset is 8640 segments of 10 seconds length, i.e. 24 hours of audio. The dataset is further partitioned into training and test subsets such that the training subset contains audio from approximately 70\% of recording locations of each city and each class. Of the total 8640 segments, 6122 segments were included in the training subset and 2518 segments in the test subset. More details on the number of segments from each location are provided in the documentation of the dataset.

\textbf{TUT Urban Acoustic Scenes Mobile 2018} development dataset contains the same recordings as TUT Urban Acoustic Scenes 2018 and, in addition, two hours of parallel data recorded with devices B and C. Therefore the dataset contains 2 hours of data recorded with all three devices (A, B and C). The amount of data is as follows:
\begin{itemize}
\item Device A: 24 hours (8640 segments, 864 segments per scene) 
\item Device B: 2 hours (720 segments, 72 segments per scene)
\item Device C: 2 hours (720 segments, 72 segments per scene)
\end{itemize}
In this dataset, the data from device A which is originally binaural, was resampled to 44.1~kHz and averaged into a single channel, to align with the properties of the data recorded with devices B and C. The dataset contains in total 28 hours of audio.

The training/test partitioning was done same as for TUT Urban Acoustic Scenes 2018, with approximately 70\% of recording locations for each city and each scene class included to the training subset,  considering only device A. The training subset contains 6122 segments from device A, 540 segments from device B, and 540 segments from device C. The test subset contains 2518 segments from device A, 180 segments from device B, and 180 segments from device C. 
The data partitioning is illustrated in Fig.~\ref{fig:datasets}.

\begin{figure}
    \centering
    \includegraphics[width=1.0\linewidth]{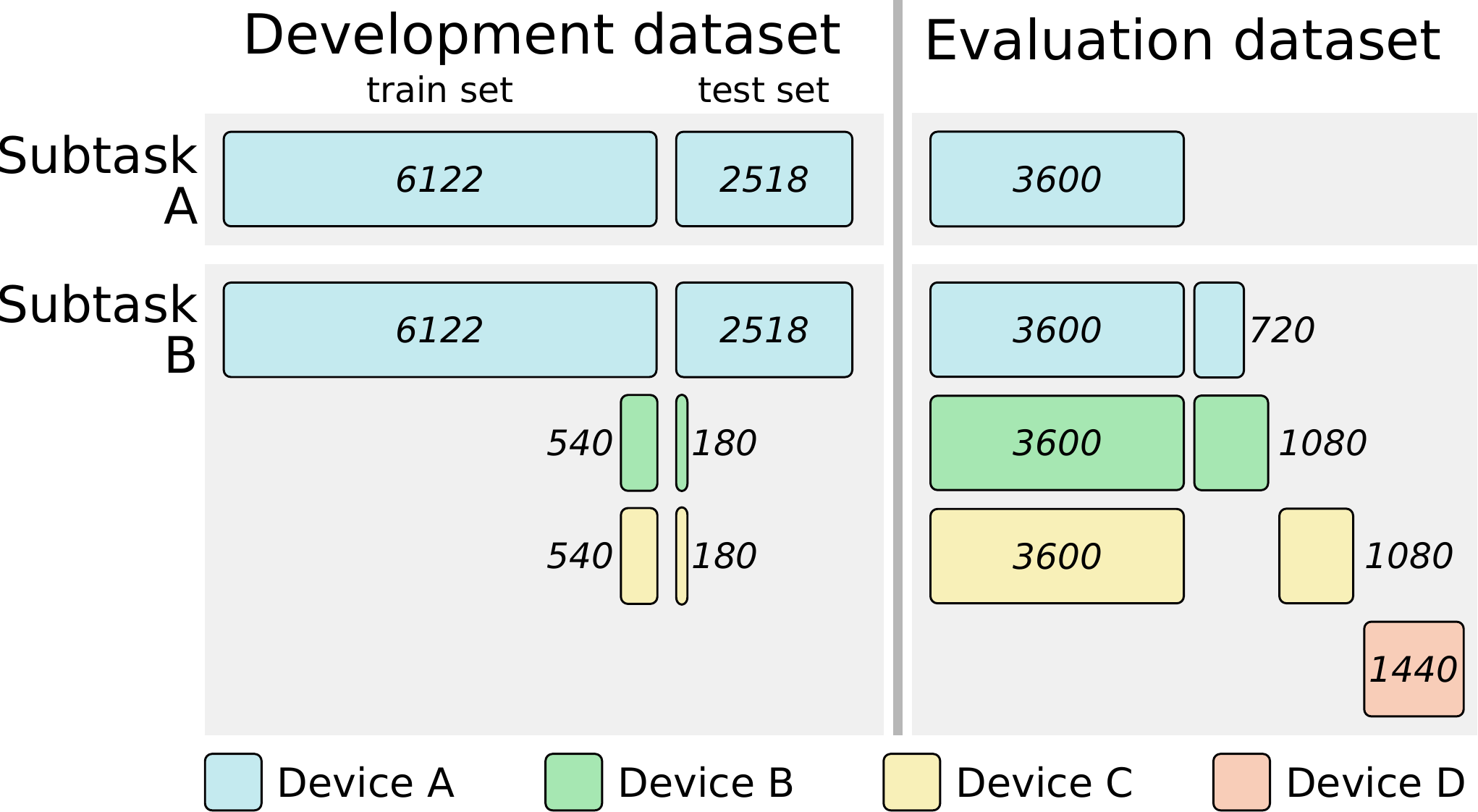}
    \caption{Development and evaluation data amounts. 
    %\newline Evaluation dataset details will be provided after the challenge submission deadline.
    }
    \label{fig:datasets}
    \vspace{-10pt}
\end{figure}

\subsection{Evaluation datasets}

\textbf{TUT Urban Acoustic Scenes 2018} evaluation dataset contains audio examples from locations different that the ones in the development dataset. 
The dataset contains 3600 segments, therefore 10 hours of audio material, all recorded with device A. In the DCASE Challenge, systems will be ranked based on classification accuracy for the evaluation dataset, calculated as class-wise average.
% info only after the deadline
% --------------------
The dataset is balanced at class level, having a number of 360 segments per scene, being as balanced as possible at city level too, with 72 segments per scene per city when possible. There are only few exceptions, notably Barcelona airport being available only in the development set.
% ----------------------------- until here 

\textbf{TUT Urban Acoustic Scenes Mobile 2018} evaluation dataset contains 42 hours of data, recorded with all four devices. The data recorded with device A was resampled and converted to single channel, just as in the development dataset. 
% info only after deadline 
% --------
The dataset contains 3600 segments of parallel data from devices A, B and C, 360 segments of non-parallel data each from devices B and C, and 1440 segments from device D. To create more diversity and prevent guessing of device specific segments, there are an additional 720 non-parallel segments from devices A, B and C, whose sole purpose is to create a non-balanced set, i.e. these segments will not be evaluated. 
% --- until here 

Ranking of the systems will be done based on classification accuracy only on audio recorded with devices B and C. Data from device A will be used for comparison with subtask A performance, while data from device D, which was not encountered at all in training, will be used to analyze performance on completely unseen devices. No information about device identity was provided with the segments, in order to force generalization and avoid tuning the systems towards the specific devices.

\section{Baseline system results}
\label{sec:results}
\vspace{-4pt}

\begin{figure}[!t]
    \centering
    \includegraphics[width=0.73\linewidth]{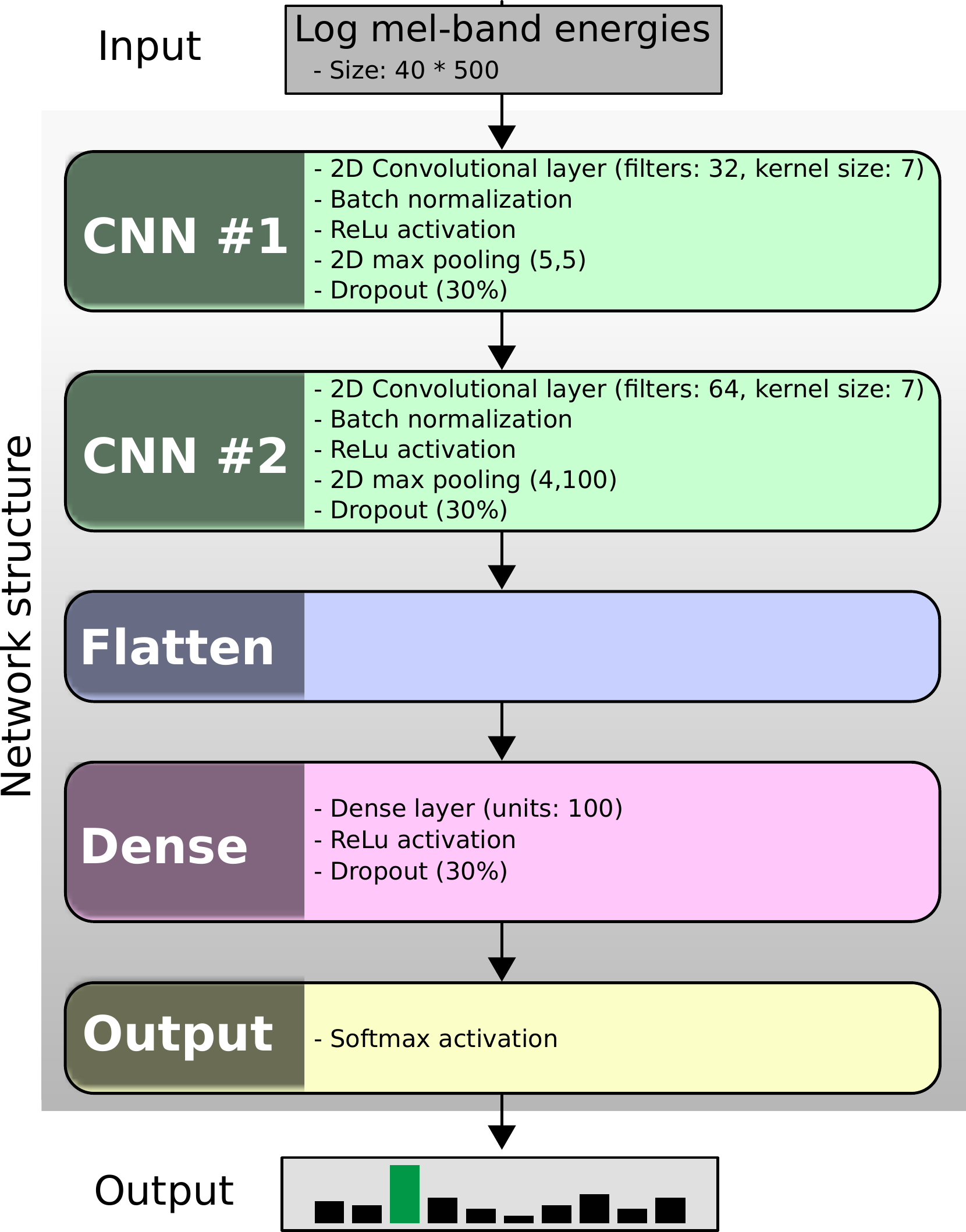}
    \vspace{-5pt}
    \caption{Baseline system architecture.}
    \label{fig:system}
    \vspace{-15pt}
    \end{figure}

\begin{table*}[!t]
\centering
\caption{Baseline system results for acoustic scene classification, subtasks A and B in DCASE 2018 challenge. In subtask B, ranking is done by average performance on data from devices B and C. 
Device ID is indicated per column for subtask B.
}
\label{tab:results}
\begin{minipage}{0.32\textwidth}\centering
\vspace{4pt}
Subtask A \\
\vspace{4pt}
\begin{tabular}{|l|cc|}
Acoustic Scene & Dev set & Eval set \\
& & \\
\hline
Airport				& 72.9 & 55.3 \\
Bus	    			& 62.9 & 66.1 \\
Metro   			& 51.2 & 60.8  \\
Metro station 		& 55.4 & 52.8 \\
Park    			& 79.1 & 79.4 \\
Public square 		& 40.4 & 33.9 \\
Shopping mall 		& 49.6 & 64.2 \\
Street, pedestrian 	& 50.0 & 55.3 \\
Street, traffic		& 80.5 & 81.9 \\
Tram 				& 55.1 & 60.0 \\
\hline
Average & \textbf{59.7}   & \textbf{61.0}  \\
& ($\pm$0.7) & \\
\end{tabular}
\end{minipage}
\begin{minipage}{0.67\textwidth}\centering
\vspace{4pt}
Subtask B \\
\vspace{4pt}
\begin{tabular}{r|ccc|c|ccc|cc|}
& \multicolumn{4}{c|}{Development set} & \multicolumn{5}{c}{Evaluation set} \\
 & B & C & avg (B,C) & A &  B & C & avg (B,C) & A & D \\
\cline{2-10}
&  68.9 & 76.1 & 72.5 & 73.4 & 65.8 & 59.4 & 62.6 & 66.8 & 1.4  \\
&  70.6 & 86.1 & 78.3 & 56.7 & 50.5 & 69.5 & 60.0 & 76.1 & 19.4 \\
&  23.9 & 17.2 & 20.6 & 46.6 & 50.2 & 40.4 & 45.3 & 61.9 & 54.2 \\
&  33.8 & 31.7 & 32.8 & 52.9 & 37.4 & 44.9 & 41.2 & 58.0 & 65.3 \\
&  67.2 & 51.1 & 59.2 & 80.8 & 58.6 & 63.0 & 60.8 & 82.9 & 6.9 \\
&  22.8 & 26.7 & 24.7 & 37.9 & 17.0 & 16.5 & 16.7 & 24.3 & 0.7 \\
&  58.3 & 63.9 & 61.1 & 46.4 & 49.2 & 55.4 & 52.3 & 62.2 & 78.5 \\
&  16.7 & 25.0 & 20.8 & 55.5 & 35.8 & 27.3 & 31.5 & 53.8 & 0.0 \\
&  69.4 & 63.3 & 66.4 & 82.5 & 69.2 & 69.9 & 69.5 & 83.1 & 25.7 \\
&  18.9 & 20.6 & 19.7 & 56.5 & 41.3 & 30.9 & 47.6 & 63.6 & 22.9 \\
\cline{2-10}
 & 45.1  & 46.2  & \textbf{45.6}  & 58.9  & 47.5 & 47.7 & \textbf{47.6} & 63.6 & 27.5 \\
 & ($\pm$3.6) & ($\pm$4.2) & ($\pm$3.6) & ($\pm$0.8) & & & & \\
\end{tabular}
\end{minipage}
\end{table*}

The baseline system implements a convolutional neural network-based approach (CNN). The architecture of the network is based on one top ranked submission from DCASE 2016  \cite{Valenti2017}, with added batch normalization and changes to layer sizes. 
This approach aims to implement a popular solution based on previous challenges, and to offer a satisfactory performance for the task. 

For each 10-second audio file, log mel-band energies were first extracted in 40 bands using an analysis frame of 40 ms with a 50\% hop size. The neural network consists of two CNN layers and one fully connected layer, and uses an input of size 40x500, equivalent to the full length of the segment to be classified. 
The network was trained using Adam optimizer \cite{kingma2014adam} with a learning rate of 0.001.
The system architecture is presented in Fig. \ref{fig:system}, including details of each layer.
The model selection was done using a validation set consisting of approximately 30\% of the original training data, selected such that training and validation sets do not have segments from the same location, and both sets have data from each city. The model performance was evaluated on the validation set after each epoch, and the best performing model was selected.

\begin{figure}[!t]
    \centering
    \includegraphics[width=0.95\linewidth]{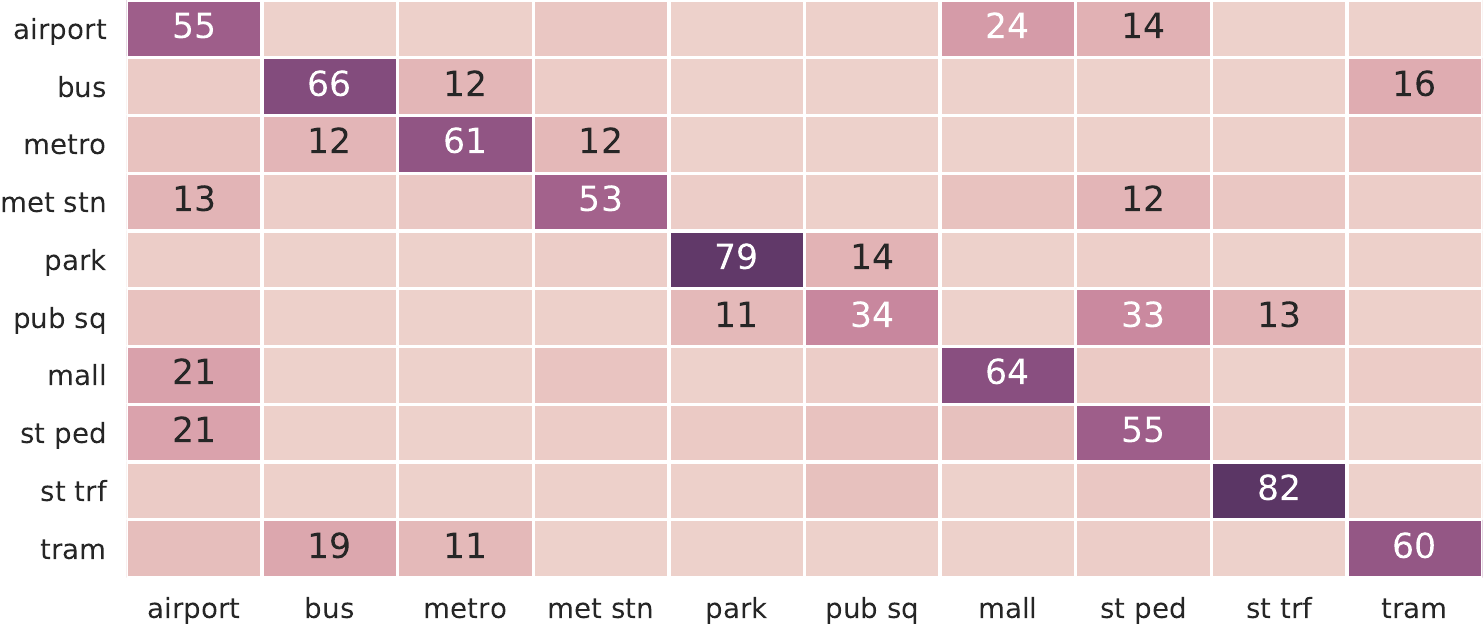}
    \vspace{-5pt}
    \caption{Confusion matrix for subtask A. Evaluation set}
    \label{fig:conf_matrix}
    \vspace{-15pt}
    \end{figure}

Table \ref{tab:results} presents the baseline system results for subtasks A and B, both on development set and evaluation sets. In the development stage, the system was trained and tested 10 times using the provided training/test split to account for the effect of random weight initialization in training; the mean and standard deviation of individual performance from these 10 independent trials is presented in the table as development set performance. 
% class-wise results, generalization
On subtask A, system performance on the development set is 59.7\%, with class-wise results varying from 40.4\% to 80.5\%. % only after challenge deadline 
% ---------------
The system generalizes rather well, having a performance of 61\% on the evaluation set. By comparing system performance on individual scene classes, we note that most scenes have very similar performance for the two sets - bus, metro station,park street with traffic. The most difficult class to recognize is public square, with the lowest performance in development set and even lower (33.9) in evaluation set, while the best performance of over 80\% is obtained for the street with traffic scene. The confusion matrix is presented in Fig.  \ref{fig:conf_matrix}. 
% ------------------ until here 

%\textcolor{red}{should we check some confusions? have a figure with confusion matrix?TV: if space allows, this will be nice to have. if it does not fit, then we should have it in some later publication}
For subtask B, the system was trained using only the audio material from device A (6122 segments of high-quality audio) to highlight the problem of mismatched recording devices. No additional techniques for dealing with the mismatch was used, in order to avoid influencing the challenge participants in their choice of method.
%TV: in some later publication we should compare how the results are if we include the training material that includes B and C also
Test subset results are presented separately for each device (2518 segments from device A, 180 segments from device B, 180 segments from device C); average of devices B and C is highlighted, as this is the official ranking measure in the challenge. Here too the system was trained and tested 10 times using the provided training/test split.

System performance on data from device A, same as its training data, is 58.9\%, comparable with the performance in subtask A. On devices B and C, the system performs similarly, with an over 10\% gap to device A, clearly showing the device mismatch. 
% only after challenge deadline 
% ---------------
The system has a similar behavior on the evaluation set, with a performance of 63.6\% for device A and 47.6\% average on devices B and C, a sign of good generalization and consistent behavior. Performance on device D is very low, with a very large gap even to devices B and C. The one important characteristic of device D is that it provides audio in compressed format, which may be the cause of such extreme mismatch.
% ----------- until here 

% only after challenge deadline 
% ---------------
Class-wise performance is mostly similar between development and evaluation sets for all devices, with the performance gap still present between devices even when the development and evaluation performance for same device is significantly different: for example metro with 20\% (devices B,C) and 46\% (device A) in development, increasing to 45\% and 61\%, respectively, in evaluation set.
% ---------------- until here 

%\textcolor{red}{might be interesting to have conf matrix of subtask A eval and subtask B eval avg(b,c) next to each other, see if distribution of confusions changes between the subtasks; if we do this we will be a bit short of space}

\section{Conclusions}
\label{sec:concl}
The acoustic scene classification within DCASE 2018 challenge offers participants three interesting subtasks, each with own research question. In subtask A, the same classification problem is approached for a dataset with a much larger size and acoustic variability than before, subtask B calls for solutions to the device mismatch problem, while subtask C allows use of external resources such as data and transfer learning for increasing classification performance. The datasets are freely available and not limited for use within the challenge, and will be extended in the future to include more cities and possibly other acoustic scene classes, to further increase task complexity through acoustic variability and allow other challenging research questions, such as training with unbalanced data, or open set classification.

\newpage

% -------------------------------------------------------------------------
% Either list references using the bibliography style file IEEEtran.bst
\bibliographystyle{IEEEtran}
\bibliography{refs}

\end{sloppy}
\end{document}